\documentclass[aps,prl,twocolumn,reprint,superscriptaddress,citeautoscript]{revtex4-2}

\usepackage{physics}
\usepackage{graphicx}
\usepackage{amsmath,amssymb}
\usepackage{bm}
\usepackage[normalem]{ulem}

\usepackage{xcolor}

\newcommand{\av}[1]{\ensuremath{\langle#1\rangle} }

\usepackage{hyperref}
\hypersetup{
	citecolor = blue,
	colorlinks = true,
	urlcolor = blue
}

\begin{document}

\title{Chirality-induced Phonon-Spin Conversion at an Interface}

\author{T. Funato}
\affiliation{Center for Spintronics Research Network, Keio University, Yokohama 223-8522, Japan}
\affiliation{Kavli Institute for Theoretical Sciences, University of Chinese Academy of Sciences, Beijing, 100190, China.}
\author{M. Matsuo}
\affiliation{Kavli Institute for Theoretical Sciences, University of Chinese Academy of Sciences, Beijing, 100190, China.}
\affiliation{CAS Center for Excellence in Topological Quantum Computation, University of Chinese Academy of Sciences, Beijing 100190, China}
\affiliation{Advanced Science Research Center, Japan Atomic Energy Agency, Tokai, 319-1195, Japan}
\affiliation{RIKEN Center for Emergent Matter Science (CEMS), Wako, Saitama 351-0198, Japan}
\author{T. Kato}
\affiliation{Institute for Solid State Physics, University of Tokyo, Kashiwa 277-8581, Japan}

\begin{abstract}
We consider spin injection driven by nonequilibrium chiral phonons from a chiral insulator into an adjacent metal.
Phonon-spin conversion arises from the coupling of the electron spin with the microrotation associated with chiral phonons. 
We derive a microscopic formula for the spin injection rate at a metal-insulator interface.
Our results clearly illustrate the microscopic origin of spin current generation by chiral phonons and may lead to a breakthrough in the development of spintronic devices without heavy elements. 
\end{abstract}

\maketitle

{\it Introduction.---}Chirality, which is defined by the breaking of the reflection and inversion symmetries of crystals, is an important concept in modern condensed-matter physics~\cite{Barron1986a,Barron1986b,Barron2012book,Barron2012,Kishine2022,Fransson2022}.
The chirality in materials has attracted much attention, in particular, after the discovery of chirality-induced spin selectivity (CISS) in DNA and peptides~\cite{Goehler2011,Xie2011,Naaman2012,Kettner2015,Michaeli2016,Naaman2019,Naaman2020,Mishra2020}.
Indeed, the discovery of CISS has stimulated a number of theoretical and experimental studies on spin-related phenomena in chiral materials~\cite{Sasao2019,Waldeck2021,Evers2022,Nabei2020,Inui2020,Shiota2021,Otsuto2021,Utsumi2022,Oiwa2022,Das2022,Suzuki2023,Nakajima2023,Togawa2023} since it may reveal a way of developing spintronic devices without using heavy elements.

The concept of chirality has been extended to the dynamical properties of solids, i.e., phonons, whose chiral nature is thought to be characterized by pseudo-angular~\cite{Boｚovic1984,Zhang-Niu2015,Tatsumi2018,Zhang2022,Komiyama2022} or angular momenta~\cite{Vonsovskii1962,McLellan1988,Zhang-Niu2014,Garanin-Chudnovsky2015,Nakane-Kohno2018,Hamada2018,Geilhufe2022,Geilhufe2023}.
Recently, the physical properties of chiral phonons have been theoretically studied~\cite{Streib2018,Juraschek2019,Chen2019,Kishine2020,Ren2021,Yao2022,Xiong2022a,Xiong2022b,Saparov2022,AKato2022,Chen2022,Bonini2023,Tsunetsugu2023} and have been experimentally observed~\cite{Zhu2018,Jeong2022,Ishito2023a,Ishito2023b,Kim2023}.
In this situation, it is natural to ask whether chiral phonons can be converted directly to electron spins at an interface or not.
However, this question remains unanswered
because of the lack of understanding of the microscopic description underlying interfacial phonon-spin conversion.

The key idea to solve this problem is reconsideration on microscopic spin-phonon coupling.
Usually, it is derived from energy change of electrons induced by lattice displacement in combination with the spin-orbit interaction.
This well-studied mechanism requires strong spin-orbit coupling, limiting its effectiveness to heavy metals and specific semiconductors. 
In our work, we consider another mechanism derived from the gyromagnetic effect~\cite{Barnett1915,Einstein-deHaas1915,Scott1962}, which has been overlooked so far. 
Previous research on the gyromagnetic effect has focused on micrometer-scale local lattice rotations, significantly larger than the lattice constant~\cite{Zolfagharkhani-NatNano-2008,Kobayashi2017,Kurimune2020,Takahashi2020,Kazerooni2020,Kazerooni2021,tateno2021Phys.Rev.B}.
Recently, Kishine et al. studied the dispersion of chiral phonons~\cite{Kishine2020} by using a degree of freedom of local rotation (so-called {\it microrotation}), within micropolar elasticity theory~\cite{Nowacki1985,eringen2012}.
However, the atomic-scale lattice rotations and their direct interactions with electron spins remain underexplored. 
This paper aims to demonstrate that the spin-microrotation coupling is the microscopic mechanism that facilitates the non-trivial interaction between chiral phonons and electron spins.

\begin{figure}
    \centering
    \includegraphics[width=80mm]{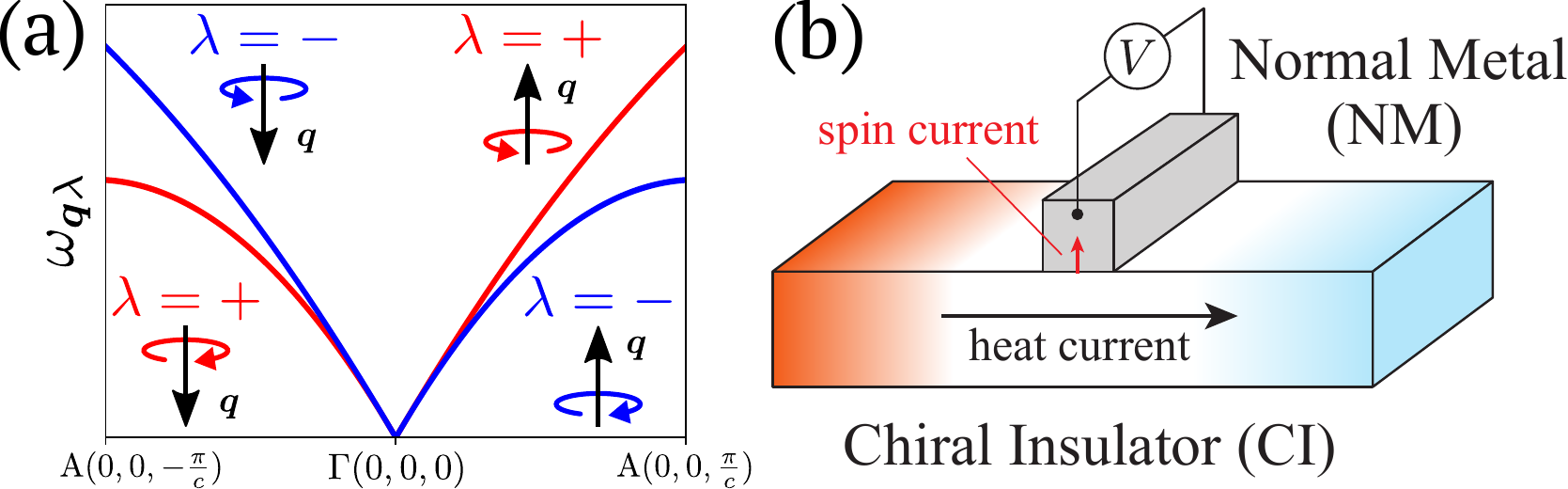}
    \caption{
    (a) Schematic illustration of energy dispersion splitting for chiral phonons.
    The red and blue lines represent the energy of the right-handed ($\lambda =+$) and left-handed ($\lambda =-$) chiral phonon modes, respectively. 
    (b) Schematic setup. Heat current in the CI generates a spin current in the NM through an interface. The generated spin current can be observed by a voltage in the NM induced by the inverse spin Hall effect.}
    \label{fig:setup}
\end{figure}

Chirality in solids is characterized by time-reversal symmetry and lack of the parity(mirror) symmetry with respect to spatial inversion.
This feature is reflected by splitting of the phonon dispersion $\omega_{{\bm q}\lambda}$ as schematically shown in Fig.~\ref{fig:setup}(a), where ${\bm q}$ is the wavenumber and $\lambda$ is the circularity of phonons.
When the phonons propagates along the chiral axis, their energy  becomes different ($\omega_{{\bm q}+}\ne\omega_{{\bm q}-}$) due to the chirality of the crystal.
The phonon dispersion lacks the parity symmetry ($\omega_{{\bm q}\lambda}\ne\omega_{-{\bm q}\lambda}$), while it keeps the time-reversal symmetry ($\omega_{{\bm q}\lambda}=\omega_{-{\bm q}\bar{\lambda}}$) where $\bar{\lambda}=\mp$ indicates the circularity opposite to $\lambda$.
We will show that this feature of chiral phonons is indeed essential to the phonon-spin conversion at an interface and directly connects with the spin current formula derived later.

The present work formulates the coupling between microrotations and electron spins (spin-microrotation coupling), thereby deriving a spin current through an interface driven by chiral phonons.
Starting with a bilayer system composed of a normal metal (NM) and chiral insulator (CI) as shown in Fig.~\ref{fig:setup}(b), we derive the effective Hamiltonian describing the interfacial coupling between the electron spins and chiral phonons due to the spin-microrotation coupling.
By treating the interfacial spin-phonon coupling perturbatively, we derive the spin current injected from the CI into the NM.
The results suggest that an imbalanced distribution among the chiral phonon modes, e.g., due to a temperature gradient, drives the interfacial spin current into the NM~\cite{ohe2024chirality}.
Our findings clearly illustrate the microscopic origin of the spin current generation by chiral phonons without the spin-orbit interaction and may lead to a breakthrough in the development of spintronic devices without heavy elements.

\begin{figure}
    \centering
    \includegraphics[width=70mm]{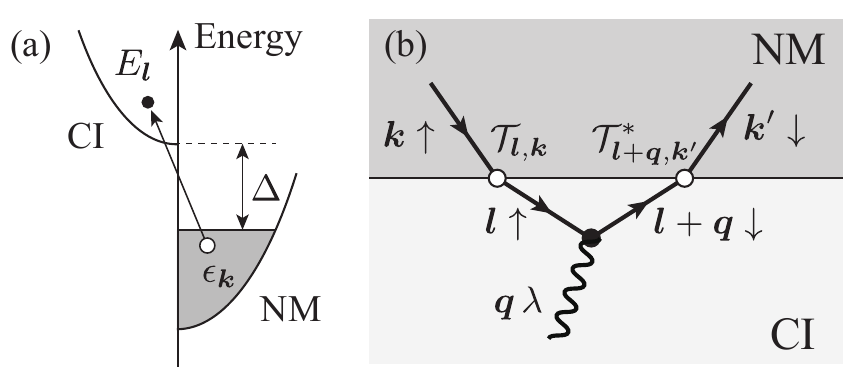}
    \caption{(a) Schematic energy diagram in the first process considered in the perturbation theory. The two parabolic curves indicate the density of states of the conduction bands in the CI and NM.
    (b) The Feynman diagram relevant to phonon-spin conversion at the interface.}
    \label{fig:diagram}
\end{figure}

{\it Model.---} We consider a NM/CI bilayer system with weak interfacial electron tunneling.
The corresponding Hamiltonian is
\begin{align}
    \hat{\mathcal H}_T = \hat{\mathcal H}_m +  \hat{\mathcal H}_e + \hat{\mathcal H}_{\text{int}} + \hat{\mathcal H}_{\text{ph}} +\hat{\mathcal H}_{\text{smc}} .
    \label{eq:total}
\end{align}
The first term $\hat{\mathcal H}_m = \sum_{\bm k\sigma}\epsilon_{\bm k} c_{\bm k\sigma}^{\dagger}c_{\bm k\sigma}$ describes the electron state in the NM, where $c^{\dagger}_{\bm k\sigma}$($c_{\bm k\sigma}$) represent the creation(annihilation) operators of electrons with eigenenergy $\epsilon_{\bm k}$ and spin $\sigma$.
The second term $\hat{\mathcal H}_e=\sum_{\bm k\sigma} E_{\bm k} d^{\dagger}_{\bm k\sigma} d_{\bm k\sigma}$ describes the electron state in the conduction band of the CI, where $E_{\bm k}$ is the eigenenergy of the conduction band and $d^{\dagger}_{\bm k\sigma}$($d_{\bm k\sigma}$) represent the creation(annihilation) operators of conduction electrons in the CI.
We assume that the conduction band of the CI is located above the Fermi energy with a large gap $\Delta$ and is empty (see Fig.~\ref{fig:diagram}(a)).
The third term $\hat{\mathcal H}_{\text{int}}=\sum_{\bm k\bm l\sigma}[\mathcal T_{\bm l,\bm k}d^{\dagger}_{\bm l\sigma} c_{\bm k\sigma} + \text{h.c.}]$ represents electron tunneling through the interface, where $\mathcal T_{\bm l,\bm k}$ is the tunneling matrix element.
The fourth term $\hat{\mathcal H}_{\text{ph}} = \sum_{\bm q\lambda} \hbar \omega_{\bm q\lambda} ( a^{\dagger}_{\bm q\lambda} a_{\bm q\lambda} + 1/2 )$ describes chiral phonons, where $\omega_{\bm q\lambda}$ and $a_{\bm q\lambda}^{\dagger}$($a_{\bm q\lambda}$) are the frequency and creation(annihilation) operator of phonons with wavenumber $\bm q$ and circularity $\lambda$, respectively.
For simplicity, we focus on acoustic modes of chiral phonons.

{\it Coupling between spin and chiral phonon.---}Next, let us derive the coupling between the phonons and electron spins represented by the fifth term $\hat{\mathcal H}_{\text{smc}}$.
The displacement of the $j$-th ion from the equilibrium position $\bm r_j$ can be expressed as
\begin{align}
    \bm u_j = \sum_{\bm q\lambda} \sqrt{\frac{\hbar}{2\rho V_{\rm CI} \omega_{\bm q\lambda}}} \bm e_{\bm q\lambda} (a_{\bm q\lambda} + a^{\dagger}_{-\bm q\bar \lambda}) e^{i\bm q\cdot \bm r_j},
\end{align}
where $\rho$ is the mass density of the lattice, $V_{\rm{CI}}$ is the volume of the CI, and $\bm e_{\bm q\lambda}$ are the polarization vectors.
There are two types of ion displacement. One is the displacement involving changes in the bond lengths and bond angles, leading to the conventional electron-phonon coupling.
The other is a local rotation characterized by the  microrotation~\cite{Kishine2020,eringen2012}, inducing a direct coupling with the electron spin.
In our study, we focus on the latter one hereafter~\footnote{We neglect usual electron-phonon coupling derived from energy change in bond lengths and bond angles, because the spin-phonon coupling derived in this way is expected to be weak for materials without heavy elements.}.
In the Born-Oppenheimer approximation,  atomic orbitals at a target atom rotate following the motion of the surrounding atoms through atomic bonds in the presence of the phonons.
Let us now introduce a corotating frame in which the atomic orbitals and the surrounding atoms do not move.
The atomic orbital at the $j$-th ion in the corotating frame, $\phi_j$, can be written as $\phi_j = \exp(i\bm J\cdot \bm \theta_j/\hbar) \phi_j^0$, where $\phi_j^0$ is the atomic orbital, in the laboratory frame, 
$\bm \theta_j$ is the rotation angle, and $\bm J=\bm L+\bm S$ is the total angular momentum consisting of the orbital and spin angular momenta, $\bm L$ and $\bm S$.
The time-dependent Schr\"odinger equation is written in the corotating frame as $i\hbar \partial_t \phi_j = (h_j - \bm J\cdot \dot{\bm \theta}_j) \phi_j$, where $h_j$ is the Hamiltonian for the atomic orbitals in the laboratory frame.
In this study, we assume that the orbital angular momentum is quenched due to crystal fields as expected in materials composed of typical elements or transition metals~\footnote{The coupling between the microrotation and the orbital angular momentum $L$ could be significant for materials including rare-earth elements or in generation of orbital currents induced by nonequilibrium driving by an external bias.
We leave discussion of these effects for future work.}. Then,
we obtain the spin-microrotation coupling, $\hat{\mathcal H}_{\theta}=-\sum_j \bm S \cdot \dot{\bm \theta}_j$.

From here on, we will restrict ourselves to long-wavelength phonons for simplicity~\footnote{The discussion here can be extended to short-wavelength phonons in principle. See Supplemental Material.}.
We note that this long-wavelength approximation is sufficient to explain spin generation at temperatures lower than the Debye temperature.
We introduce a continuous lattice displacement constituted by smoothly connecting the discretely located ions, i.e., $\bm u(\bm r) = \sum_{\bm q} u_{\bm q} e^{i\bm q\cdot \bm r}$, which is governed by the elastic equations.
In the low-frequency region, the microrotations of the ions, which is characterized by the vorticity $\bm \Omega(\bm r)=\nabla \times \dot{\bm u}(\bm r)$, adiabatically interlock with one another. 
In this case, the angle of the  microrotation and the vorticity can be related thorugh $\dot{\bm \theta}_j = \bm \Omega (\bm r_j)/2$.
Therefore, the spin-microrotation coupling reduces to a second-quantized expression of the spin-vorticity coupling:
\begin{align}
    \hat{\mathcal H}_{\text{smc}} = -\frac{\hbar}{2} \sum_{\bm q\lambda} \hat{\bm \sigma}_{-\bm q} \cdot \bm \Omega_{\bm q\lambda},
\end{align}
where $\hat{\bm \sigma}_{-\bm q} = (1/2)\sum_{\bm l\sigma \sigma'} d^{\dagger}_{\bm l+\bm q\sigma} \bm \sigma_{\sigma \sigma'} d_{\bm l\sigma'}$ is the spin density operator of the electrons in the CI with $\bm \sigma=(\sigma^x,\sigma^y,\sigma^z)$ being the Pauli matrices acting on the spin space.
Here, $\bm \Omega_{\bm q\lambda}$ is the Fourier component of the vorticity, whose second-quantized expression is given by
\begin{align}
    \bm \Omega_{\bm q\lambda} = \sqrt{\frac{\hbar \omega_{\bm q\lambda}}{2\rho V_{\rm{CI}}} } (\bm q\times \bm e_{\bm q\lambda})
\qty( a_{\bm q\lambda} -
a_{-\bm q\bar \lambda}^{\dagger} ) ,
\end{align}
and $\bm \Omega_{-\bm q\bar \lambda}^{\dagger} =\bm \Omega_{\bm q\lambda}$ holds. 

It should be noted that the vorticity is connected with the dispersion of chiral phonons through the micromotion.
Recently, Kishine et al. revealed that microrotation is vital to demonstrating the features of the chiral phonon dispersion~\cite{Kishine2020}. 
The spin-vorticity coupling derived here constitutes a coupling inherently tied to the microscopic properties of chiral phonons, unlike the phonon angular momentum that has been previously discussed~\cite{Zhang-Niu2014,Garanin-Chudnovsky2015,Nakane-Kohno2018,Hamada2018}. 
The spin-vorticity coupling, together with its origin, the spin-microrotation coupling, can be regarded as fundamental interactions between chiral phonons and electron spins. 
They potentially play a crucial role in phonon-spin conversion processes linked to chirality in various systems, from bulk materials to junctions. 

{\it Interfacial spin-phonon coupling.}---First, let us derive the effective Hamiltonian describing the conversion between the chiral phonon and electron spins.
We define the projection operator $\mathcal P$ to restrict the model space composed of the electron system in the NM and the phonon system in the CI.
The effective Hamiltonian up to second order in the electron tunneling process and first order in the spin-microrotation coupling is
\begin{align}
	\hat{\mathcal H}_{\text{e-ph}} = \mathcal P \hat{\mathcal H}_{\text{int}} \frac{1}{E_0 - \hat{\mathcal H}_0} \mathcal Q \hat{\mathcal H}_{\text{smc}} \frac{1}{E_0 - \hat{\mathcal H}_0} \mathcal Q \hat{\mathcal H}_{\text{int}} \mathcal P,
\end{align}
where $\hat{\mathcal H}_0=\hat{\mathcal H}_m +\hat{\mathcal H}_e+ \hat{\mathcal H}_{\text{ph}}$ is the unperturbed Hamiltonian, and $E_0$ is the unperturbed energy.
The operator $\mathcal Q = 1-\mathcal P$ describes the projection of the complementary space to the model space.
The corresponding Feynman diagram is illustrated in Fig.~\ref{fig:diagram}(b).
By a straightforward calculation, the effective Hamiltonian can be given the form of an interfacial coupling between the electron spin of the NM and phonon vorticity of the CI as
\begin{align}
    \hat{\mathcal H}_{\text{e-ph}} = -  \sum_{\bm p\bm q\lambda} J_{\bm q,\bm p} \qty( \Omega^+_{\bm q\lambda} \hat s^-_{-\bm p} + \Omega_{-\bm q\bar \lambda}^- \hat s_{\bm p}^+ ),
\end{align}
where $\Omega^{\pm}_{\bm q\lambda} = \Omega^x_{\bm q\lambda} \pm i\Omega_{\bm q\lambda}^y$ is the ladder operator of the vorticity, and $\hat s^{\pm}_{\bm p} = \hat s^x_{\bm p} \pm i\hat s^y_{\bm p}$ is that of the electron spin with $\hat s^{\alpha}_{\bm p}=(1/2)\sum_{\bm k\sigma \bar \sigma}c^{\dagger}_{\bm k-\bm p\sigma} \sigma^{\alpha}_{\sigma \bar \sigma} c_{\bm k\bar \sigma}$  denoting the spin density operator in the NM.
Here, $J_{\bm q,\bm p}$ is the matrix element of the interfacial spin-phonon coupling (detailed expression is given in Supplemental Material~\cite{Supplement}).
In this study, we have assumed that the electron tunneling at the interface is local and its wavenumber dependence is negligible~\cite{mahan2000many}

{\it Interfacial spin current.---}The spin current from the CI into the NM can be calculated with the effective Hamiltonian $\hat{\mathcal H} = \hat{\mathcal H}_m + \hat{\mathcal H}_{\text{ph}} + \hat{\mathcal H}_{\text{e-ph}}$.
The interfacial spin current operator is defined by the change in the total spin in the NM per unit time:
\begin{align}
\hat I_s &\equiv -\hbar \pdv{\hat s^z_{\bm 0}}{t} = i [ \hat s^z_{\bm 0} , \hat{\mathcal H}] \nonumber \\
&=i\sum_{\bm p\bm q\lambda} J_{\bm q,\bm p} \qty( \Omega^+_{\bm q\lambda} \hat s^-_{-\bm p} - \Omega^-_{-\bm q\bar \lambda} \hat s^+_{\bm p} ).
\end{align}
Therefore, the average of the interfacial spin current is given as
\begin{align}
\av{\hat I_s(t)}= \Re \qty[2i\sum_{\bm p \bm q \lambda} J_{\bm q,\bm p} \av{\hat s_{-\bm p}^-(t) \Omega^+_{\bm q\lambda}(t)}],
\end{align}
where $\hat{s}^-_{\bm p}(t)$ and $\Omega^+_{\bm q\lambda}(t)$ indicate the interaction representation, and the average $\av{\cdots}$ is taken for the Hamiltonian $\hat{\mathcal H}$.
Let us consider a second-order perturbation with respect to $\hat{\mathcal H}_{\text{e-ph}}$~\footnote{This calculation of the second-order perturbation is common as the one used to derive the current formula for tunnel junctions.}.
Here, the statistical average of the interfacial spin current becomes~\cite{Supplement,Bruus2004}
\begin{align}
&\av{\hat I_s} = \frac{\hbar^2 }{\rho V_{\rm{CI}}} \int_C d\tau \sum_{\bm p\bm q\lambda} |J_{\bm q,\bm p}|^2 \omega_{\bm q\lambda} (\bm q\times \bm e_{\bm q\lambda})_+ (\bm q\times \bm e^*_{\bm q\lambda})_- 
\nonumber \\
&\times \Re \biggl\{ \chi_{\bm p}(\tau -\tau_2) 
	\Bigl[ 
	   \mathcal D_{\bm q\lambda}(\tau_1-\tau) + \mathcal D_{-\bm q\bar \lambda}(\tau-\tau_1)
	\Bigr]
\biggr\},
\label{eq:Isinteg}
\end{align}
where the integral is taken over the Keldysh time $\tau = (t',\eta)$ with $\eta = \pm$ and the two time variables, $\tau_1 = (t,+)$ and $\tau_2 = (t,-)$, are respectively put on the forward and backward branches of the Keldysh contour.
The functions $\mathcal D_{\bm q\lambda}(\tau)= -(i/\hbar) \av{T_Ka_{\bm q\lambda}(\tau)a_{\bm q\lambda}^{\dagger}}_0$ and $\chi_{\bm p}(\tau)=(i/\hbar)\av{T_K \hat s^+_{\bm p} (\tau) \hat s^-_{-\bm p}}_0$ are respectively the phonon Green function and spin susceptibility in Keldysh form, where the average $\av{\cdots}_0$ is taken for the unperturbed Hamiltonian and $T_K$ is the time-ordering operator on Keldysh contour $C$.
These Keldysh Green functions (response functions) include four components, i.e., the lesser, greater, retarded, and advanced components~\cite{Rammer1986,Bruus2004,Stefanucci2013}.
Using the lesser and retarded components, the nonequilibrium distribution functions are defined for spin excitations in the NM and chiral phonons in the CI as
\begin{align}
f^m_{\bm p}(\omega) &= \chi^<_{\bm p}(\omega)/2i\Im \chi^R_{\bm p}(\omega), \\
f^{\text{ph}}_{\bm q,\lambda}(\omega) &= \mathcal D^<_{\bm q\lambda}(\omega)/2i\Im \mathcal D^R_{\bm q\lambda}(\omega). 
\end{align}
Here, the superscripts, $<$ and $R$, indicate the lesser and retarded components.
We assume that the NM remains in thermal equilibrium due to its high thermal conductivity and the distribution function $f^m_{\bm p}(\omega)$ coincides with the Bose-Einstein distribution $f_0(\omega,T_m) = (e^{\hbar \omega/k_BT_m}-1)^{-1}$ with temperature $T_m$.
By simplifying the convolution integral with the Fourier transformation, the spin current is calculated as~\cite{Supplement}
\begin{widetext}
\begin{align}
    \av{\hat I_s^{\alpha}} =
\frac{4\hbar^2 |J|^2}{\rho V_{\rm{CI}}} \sum_{\bm p \bm q \lambda} \omega_{\bm q\lambda}
(\bm q \cdot \hat \alpha ) 
\qty[\bm q \cdot \Im \qty(\bm e_{\bm q\lambda}^* \times \bm e_{\bm q\lambda})] 
\int^{\infty}_{-\infty} \frac{d\omega}{2\pi}
\text{Im}\, \chi^R_{\bm p} (\omega) \Bigl[-\Im \mathcal D^R_{\bm q\lambda}(\omega)\Bigr] 
\Bigl[
f^{\text{ph}}_{\bm q,\lambda}(\omega)
-
f_0(\omega,T_m)
\Bigr] . \label{eq:Formula}
\end{align}
\end{widetext}
where the formula is extended to include the spin current with polarization in $\hat \alpha$-direction ($\hat \alpha$: a unit vector indicating the spin direction to be measured).
For a rough interface, the random average of the matrix elements $|J_{\bm q,\bm p}|^2$ reduce to $|J|^2 = (\hbar |\mathcal T|^2/4\Delta^2)^2 N_b/N_N^2$, where $N_b$ is the bond number at the interface, $N_N$ is the number of the unit cells in the NM, and $|\mathcal T|$ represents the magnitude of the interfacial electron tunneling~\cite{Supplement}. 
The formula (\ref{eq:Formula}) shows that the spin current is generated when the phonon distribution function $f_{{\bm q},\lambda}^{\rm ph}(\omega)$ in the CI is driven away from the thermal equilibrium distribution $f_0(\omega, T_m)$.
We emphasize that in the formula (\ref{eq:Formula}), the chirality of the material is reflected by the factor $(\bm q \cdot \hat \alpha) [\bm q \cdot \Im (\bm e_{\bm q\lambda}^* \times \bm e_{\bm q\lambda})]$~\footnote{The spin current is proportional to both the spin spectral weight ${\rm Im} \, \chi^R_{\bm p} (\omega)/\pi$ and the density of states of chiral phonons $- {\rm Im} \, \mathcal{D}^R_{\bm q\lambda}(\omega)/\pi$.
We note that this form is analogous to the current formula for metallic tunnel junctions~\cite{Bruus2004}.}.
Actually, the axial vector $\bm e^*_{\bm q\lambda} \times \bm e_{\bm q\lambda}$ points to the chiral axis, along which the asymmetry under the mirror transformation exists.
The flipping of structural chirality, i.e., a mirror operation on a plane perpendicular to the chiral axis, inverts the polarization of the chiral phonons, and consequently, reverses the flowing direction of the spin current through the factor $\bm e^*_{\bm q\lambda} \times \bm e_{\bm q\lambda}$.
Furthermore, if the material has symmetry with respect to the mirror operation, the spin current becomes zero after a summation with respect to ${\bm q}$.
This is our main result.

{\it Temperature gradient.}---Let us derive a formula for the spin current injected into the NM due to steady phonons flows driven by a temperature gradient (see Fig.~\ref{fig:setup}(b)).
We assume that the temperature gradient is in the $z$-direction (the same as the chiral axis) and is of much larger scale than the typical mean-free path of phonons $l_{\text{ph}}$, i.e., $l_{\text{ph}} |\partial_z T/T| \ll 1$.
The steady-state phonon distribution is calculated from the Boltzmann equation as $\bm v_{\bm q\lambda} \cdot \nabla f^{\text{ph}}_{\bm q,\lambda} = (f^{\text{ph}}_{\bm q,\lambda}-f_0)/\tau_{\bm q\lambda}$, where we have employed the relaxation-time approximation, and $\bm v_{\bm q\lambda}=\partial_{\bm q} \omega_{\bm q\lambda}$ is the velocity of phonons, and $\tau_{\bm q\lambda}$ is the momentum relaxation time of phonons, respectively.
Solving the Boltzmann equation up to first order in the temperature gradient, the nonequilibrium part of the phonon distribution function is found to be $\delta f^{\text{ph}}_{\bm q,\lambda} =f^{\text{ph}}_{\bm q,\lambda} -f_0= \tau_{\bm q\lambda} (\hbar \omega_{\bm q\lambda}/k_BT) v_{\bm q\lambda}^z \qty( -\partial_z T/T )$.

We can proceed in calculation using $\Im \mathcal D^R_{\bm q\lambda}(\omega) = -\pi \delta (\hbar \omega - \hbar \omega_{\bm q\lambda})$ and $\sum_{\bm p} \Im \chi^R_{\bm p}(\omega) = \pi \nu_F^2 N_N^2 \hbar \omega$, where $\nu_F$ is the density of states at the Fermi level per unit cell in the NM.
For simplicity, we roughly approximate the polarization vector as $\bm e_{\bm q\pm} = (\hat x\pm i\hat y)/\sqrt 2$~\footnote{Although we should consider the detailed wavenumber dependence of the polarization vector for accurate calculation, such an effect only changes the prefactor of the spin current.}, and assume wavenumber-independent relaxation time as $\tau_{\bm q\lambda} = \tau$.
Thus, the spin current is calculated as
\begin{align}
    \av{\hat I_s^z} = 
     \frac{\pi \tau \hbar^3 \nu_F^2 N_N^2 |J|^2}{\rho V_{\rm{CI}} k_BT}
    \sum_{\bm q \in q_z>0} q^2_z \pdv{q_z} \qty( \omega_{\bm q+}^4 - \omega_{-{\bm q} +}^4 ) \qty( -\frac{\partial_z T}{T} ),
    \label{eq:thermal}
\end{align}
where the sum has been restricted into positive $q_z$ using the time-reversal symmetry $\omega_{-\bm q\lambda}=\omega_{\bm q \bar\lambda}$ and the symmetry of the distribution function, $\delta f^{\text{ph}}_{-\bm q,\bar \lambda}= -\delta f^{\text{ph}}_{\bm q,\lambda}$, under temperature gradient.
This result indicates that the temperature gradient along the chiral axis in the CI generates spin current into the NM across the junction.
As clearly shown from Eq.~(\ref{eq:thermal}), this spin current is generated only when the material has the structural chirality, i.e., lacks the parity symmetry ($\omega_{\bm q\lambda}\ne \omega_{-\bm q \lambda}$)~\footnote{The nonequilibrium distribution function of chiral phonon is essential to the present spin-current generation.
Such a nonequilibrium condition can be achieved by other methods, e.g., by dynamic excitation of phonons by light.}.

Finally, let us estimate the spin current generated by temperature gradient.
For simplicity, we use the dispersion relation of chiral phonons calculated in Ref.~\cite{ishito2023chiral} as a typical example.
We set the parameters as $\tau = 10^{-10}\,\text s$~\cite{ishito2023chiral}, the lattice constant $c=1\,\text{\AA}$, the mass of the unit cell as $M=\rho c^3 =10^{-26}\,\text{kg}$, $\nu_F = 10^{-2}\,\text{eV}^{-1}$, and $|\mathcal T|/\Delta=1/20$. 
By assuming a linear temperature gradient of 1\% at $1\, \text{mm}$ length, the spin current is estimated for a $10^4\,\mu \text m^2$ junction as $\av{\hat I^z_s} \sim 100\, \text{nA}$, which is observable in the present experimental technique.
After our submission, we encountered an experimental study that observed a thermal-gradient-induced spin current in a bilayer composed of $\alpha$-quartz and tungsten~\cite{ohe2024chirality}. Our findings are in good agreement with this experiment.

{\it Discussion.}---In this study, our primary focus has been on the long-wavelength acoustic phonon modes, where the continuum approximation can be applied to relate vorticity with microrotation.
However, above the Debye temperature, the consideration of optical modes (or short-wavelength phonons) becomes necessary.
We stress that spin-microrotation coupling can be established also for optical modes through careful examination of the rotation of atomic orbitals, induced by the motion of surrounding atoms. This expansion of our study will facilitate the evaluation of spin-microrotation coupling for specific materials, utilizing energy dispersion and lattice displacement obtained from first-principles calculations. Further elucidation of this extended framework will be presented in a separate publication.

{\it Summary.}---We investigated spin injection into a metal driven by nonequilibrium chiral phonons in an adjacent insulator. We constructed an effective Hamiltonian to depict the spin-phonon conversion at an interface attributable to the spin-microrotation coupling and computed the spin current considering the nonequilibrium distribution function of chiral phonons.
The results of our study imply that the spin current at the interface into a non-magnetic material is caused by an imbalanced distribution of phonons along the chiral axis. 
Importantly, our findings present a solid groundwork for chirality-governed spintronics via phonons, bypassing the requirement for a strong spin-orbit interaction with heavy elements. Future studies will apply these findings to spin-induced phenomena arising from chiral phonons.

\begin{acknowledgments}
We would like to thank J. Kishine, H. Nakayama, T. Horaguchi, and Y. Nozaki for their valuable and informative discussions.
We also thank Y. Togawa and M. Kato for the seminar talks which provided us a chance to consider spin generation by chiral phonons.
This work was partially supported by JST CREST Grant No.~JPMJCR19J4, Japan.
We acknowledge JSPS KAKENHI for Grants (No.~JP20K03831, No.~JP21H04565, No.~JP21H01800, No.~JP21H04565, and No.~JP23H01839). 
\end{acknowledgments}


%

\end{document}